**Influence of aqueous electrolytes on electrochemical performance of vertical graphene nanosheets supercapacitor electrode**

Subrata Ghosh* [1,2], Bhavana Gupta † [2,3], Tom Mathews [1,2], A. Das [1,2] and M. Kamruddin * [1,2]

[1]Homi Bhaba National Institute, Mumbai – 400094, India;

[2]Surface and Nanoscience Division, Materials Science Group, Indira Gandhi Centre for Atomic Research, Kalpakkam - 603102, India;

[3]Department of Solar Energy and Environmental Physics, Ben-Gurion University of the Negev, Sede Boqer Campus, Midreshet Ben-Gurion 84990, Israel.

## Abstract

Vertical graphene nanosheets (VGN) grown as controlled porous network are studied and demonstrated as a promising electrode material for supercapacitors. The VGN synthesized by microwave plasma enhanced chemical vapor deposition using $CH_4$/Ar gas mixture as precursor are considered for electrochemical performance in $Na_2SO_4$, KOH, and $H_2SO_4$ to delineate the electrolyte effect. Among the electrolytes, $H_2SO_4$ exhibited excellent specific areal capacitance (188 $\mu F/cm^2$) and good capacitance retention (96.8%). No significant change is observed in impedance spectra even after 200 cycles. An electric equivalent circuit for the system is simulated from Nyquist plot to elucidate the behavior of electrode/electrolyte interface. This potential supercapacitor electrode material is well characterized by scanning electron microscopy, transmission electron microscopy, Raman spectroscopy, and contact angle measurement. Utilization of aqueous electrolytes for potential supercapacitors is also discussed in relation to improved performance observed in $H_2SO_4$ medium.

*Subrata Ghosh (subrataghosh.phys@gmail.com) and *M. Kamruddin (kams@igcar.gov.in)

## 1. Introduction

Electric double layer capacitors (EDLC) or supercapacitors (SCs) have become the most promising candidates for energy storage devices including portable devices, hybrid electric vehicles, digital telecommunication system and other renewable energy storage applications due to its high power density (>10 kW/kg) and impressive cycle life (>$10^5$cycles).[21] However, the low energy density of SCs compared to that of conventional batteries limits its commercialization. The energy density can be improved by choosing electrolytes with higher potential window and higher dielectric constant as well as by architecturing the electrode material with high surface area and optimized pore distribution. Off late, considerable research efforts have been devoted in improving the energy density without sacrificing the power density and cycle life by choosing advanced material with high specific capacitance and suitable electrolyte. Porous carbon nanomaterials, such as activated carbon, carbide derived carbons, mesoporous template carbon, carbon nanotube (CNT), graphene and their derivatives, are being extensively studied as promising materials because of their high surface area, good electrical conductivity, chemical stability, cost-effectiveness and bio-degradability.[4, 9, 25] The charge storage mechanism of porous carbon based SCs is formation of electric double layer (EDL) at the electrode/electrolyte interface, without faradic reactions. However, use of binder, restacking and agglomeration during electrode fabrication lowers the effective surface area and increase the internal resistance thereby limiting the power density. To alleviate the drawback, a great deal of research has been spent on synthesizing vertically oriented porous carbon nanostructures. These structures have more surface area, channels and multiple pathways for electrolyte ion movement. In addition, presence of large amount of edge planes on electrode surface enhances the capacitance value.[1, 22] In view of this, Kim et al. have grown CNTs on graphene and studied the electrochemical performance. In this system, graphene serve as the current collector and CNTs with high surface area as EDL former.[18] Capping of CNTs, by the catalyst used during growth, renders tube inner surface and top edge unavailable for charge storage. Hence, development of a catalyst-free high surface area carbon based material which can act both as electrode and current collector is very important.

The vertical graphene nanosheets (VGN), because of their controlled porous network, high surface area, high thermal and chemical stability, excellent electronic conductivity and bio-degradability, is a highly promising SC electrode material. Therefore, the last few years have witnessed a strong momentum towards the electrochemical characterization of VGN for application in SCs.[1, 6, 7, 19, 22, 26, 28, 30, 36, 38]

Table 1: comparison on capacitance values of VGNs under different aqueous electrolytes reported in the literature.

| Ref. | height | substrate | Aqueous electrolyte | Sp. capacitance |
|---|---|---|---|---|
| **[22]** | 0.6 μm | Ni | 25 wt % KOH | 175 μF/cm$^2$at 120 Hz |
| **[26]** | 0.1μm | Pt/Si | 6M KOH | 704 μF/cm$^2$ at 2 mV/s |
| **[38]** | - | carbon paper | 6M $H_2SO_4$ | 76 mF/cm$^2$ |
| **[36]** | 0.5 μm | Carbon cloth | 1M $H_2SO_4$ | 0.7 F/cm$^2$ at 2 mV/s |
| **[28]** | - | Ni foam | 1M $Na_2SO_4$ | 0.46 F/cm$^2$at 5 mV/s |
| **[19]** | - | Glassy carbon | 1M $Na_2SO_4$ | 0.8-1 mF/cm$^2$ |

The overall SC performance not only depends on the electrode material but also on the electrolyte used.[17, 20, 34] The use of aqueous electrolyte for SC application is more suitable than the organic electrolytes in terms of cost, life time, low internal resistance, non-flammability, and toxicity. The different VGN/aqueous electrolyte systems studied are listed in **table 1**. Since, compatibility of ion-size with pore-size, charge transfer resistance and ionic mobility of electrolytes are major driving parameters for better performance, a good understanding of electrode-electrolyte interaction is essential.

In the present study, the electrochemical performance VGN/$Na_2SO_4$, VGN/KOH and VGN/$H_2SO_4$ systems are investigated by cyclic voltammetry (CV), charge-discharge and electrochemical impedance spectroscopy (EIS). An electrical equivalent circuit is simulated from the impedance plot to understand the VGN/aqueous electrolyte interfaces.

## 2. Experimental methods

### 2.1. Synthesis

The VGN are directly grown on n-Si (100) by electron cyclotron resonance chemical vapor deposition (ECR-CVD) technique. The growth and system details are published elsewhere.[10, 12] However, the procedure in brief is as follows: after cleaning the substrate in acetone, isopropyl alcohol and de-ionized water, it is dried in flowing dry $N_2$. It is then loaded into the chamber and evacuated to $2\times10^{-5}$ mbar using a turbomolecular pump. The substrate is then cleaned using Ar plasma, at 800°C, to remove the organic residue and create localized hot spots. These localized hot spots are energetically favored to adsorb the hydrocarbon species. The VGN are then grown on the Si(100) substrates maintained at 800°C by striking plasma under 5 sccm flow of $CH_4$ and 25 sccm flow of Ar at $10^{-3}$ mbar operating pressure. The microwave power is maintained at 400W. After 1 hr of growth, the gas flow and plasma are switched off and system is allowed to cool down to room temperature. The sample is then taken out for characterization and electrochemical investigation.

**2.2. Characterization**

The morphology of the films is investigated by field emission scanning electron microscope (FE-SEM; Supra 55, Zeiss) and high resolution transmission electron microscope (HRTEM; LIBRA 200FE, Zeiss). The samples for HRTEM are prepared by scratching VGN from the Si substrates and placing it on the Cu grids coated with lacey carbon. The Raman scattering measurements are carried out using a micro-Raman spectrometer (Renishaw in-Via, UK) to evaluate the structural properties in terms of crystallinity. In backscattering geometry, the spectra are taken in the 1100-3500 $cm^{-1}$ region for 30s using 514.5 nm laser and 100× objective lens. The Raman spectra are fitted with Lorentzian lineshape by WIRE 3.2 software. The sheet resistance of VGN is measured using Agilent B2902A precision source/measure unit by making four contacts using conductive silver paste. The contact angle of the VGN surface with water, 1M $Na_2SO_4$, 1M KOH and 1M $H_2SO_4$ are measured by sessile drop method with the help of a CCD camera (Apex Instrument Co. Pvt. Ltd., India). The volume of the droplet is about 2µl and all measurements are carried out in ambient conditions. The value of contact angle is evaluated by half angle fitting method provided with the instrument.

The electrochemical properties of the VGN electrode are studied using a three-electrode system in 1M $Na_2SO_4$, 1M KOH, and 1M $H_2SO_4$ electrolyte media. All the chemicals are of analytical grade and used without any further purification. The VGN, Ag/AgCl (3M KCl) and Pt foil (1.5 cm × 1.5 cm) are used as working, reference and counter electrodes, respectively. The electrical contact on top of VGN is made by thin copper wire and Ag paste. Cyclic voltammetry, galvanometric charge-discharge and EIS studies are performed using Metrohm-Autolab electrochemical work station (model PGSTAT302N, Netherland). The CV behaviors are carried out in a potential window of 0.5 V vs Ag/AgCl (3M KCl saturated) at scan rates ranging from 100 to 500 mV/s. The EIS measurements are performed in the frequency range 10 kHz to 0.1 Hz at open circuit potential with an ac perturbation of 10 mV. The Nyquist plots are drawn using the impedance data and fitted with appropriate equivalent electric circuit using EIS spectra analyzer software.[5]

## 3. Results and discussion

### 3.1. Morphological analysis

The SEM micrographs of VGN depicted in Figure 1(a-b), reveal that the vertical sheets are very sharp and the sheet length vary from ~ 100 nm to a micron. As it can be seen from Figure 1(a-b), the film consists of interconnected three-dimensional network of vertical sheets and the gap between the walls varies from 40 to 250 nm. The open macroporous structure can serve as "ion buffering reservoir" such that the electrolyte ions can access full sheet surface area.[33] Thus each vertical sheet can serve as a nanoelectrode.[14] The micropore and mesopore structure nested inside the macroporous structure, can be seen from the micrographs (Figure 1 (a-b)). This type of morphology is very useful for charge accommodation. As a result, VGN are expected to have better capacitative performance during electrochemical charge-discharge process. The cross sectional image shown in Figure 1(c) reveals that the VGN are folded, standing perpendicularly to the substrates and are of 470 nm height. The panels (d) and (e) of Figure 1 depict the TEM and HRTEM images of VGN. Figure 1(e) confirms that each vertical sheet consists of 2-10 layers of graphene. The interlayer spacing is found to be 0.36 nm, confirming its turbostratic nature.

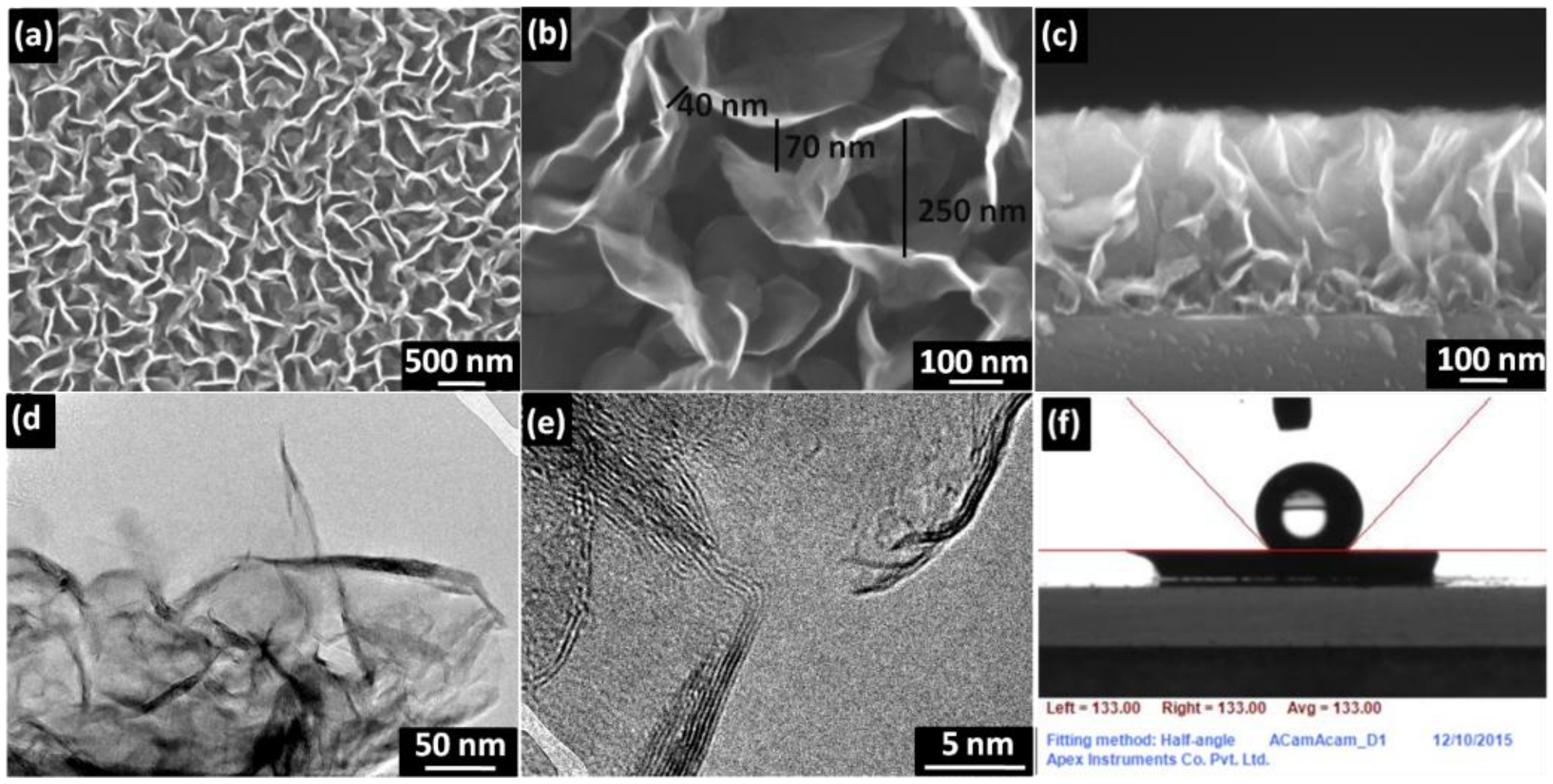

Figure 1: (a-b) Top view and (c) cross sectional view of SEM micrographs; (d) TEM and (e) HRTEM micrograph; and (f) water contact angle of vertical graphene nanosheets (VGN).

The wetting property of the film is one of the key parameters for application as SC electrode. Figure 1(f) confirms the near superhydrophobic nature of the film with water contact angle of 133°. The initial contact angles of VGN with 1M $Na_2SO_4$, 1M KOH and 1M $H_2SO_4$ electrolyte solution are 126.8°, 118.9° and 130.7° respectively. The time dependent wettability of VGN with the electrolyte solutions are shown in Figure 2. The slope of the curve represents change in wettability (hydrophobic to hydrophilic) with time and found to be 0.74°, 1.77° and 0.81° per minute. The result clearly indicates that the wettability of KOH is higher than the other studied electrolytes.

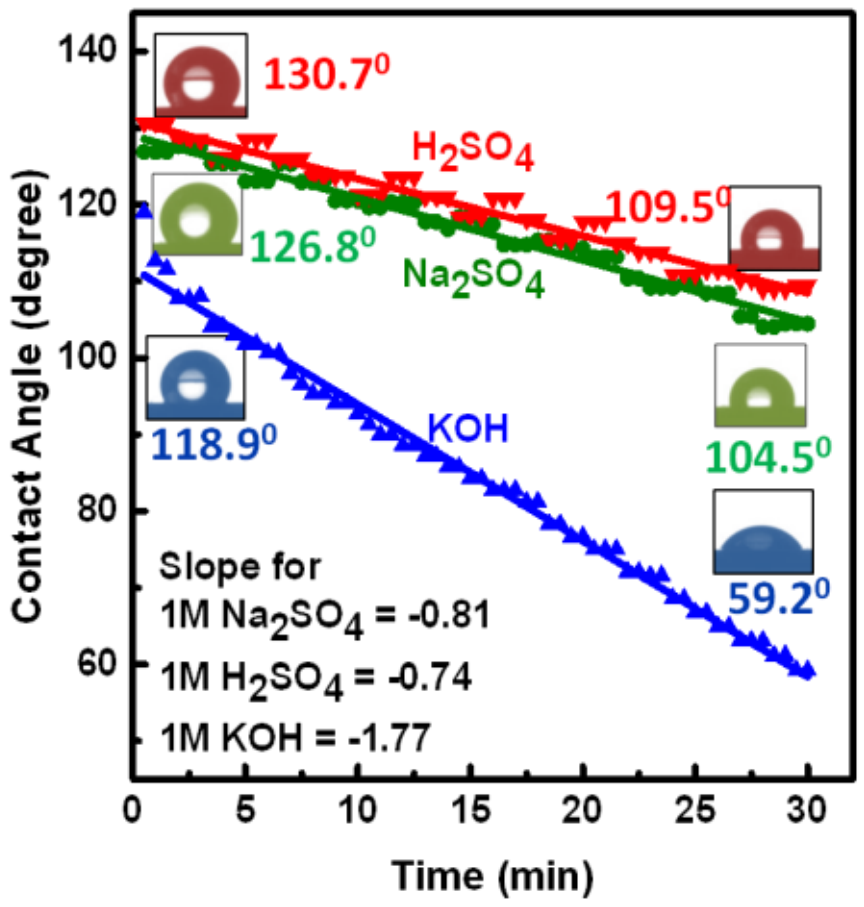

Figure 2: Time dependent wettability of vertical graphene nanosheets in1M $Na_2SO_4$, 1M KOH, and 1M $H_2SO_4$ electrolyte medium.

### 3.2. Raman spectra analysis

The crystalline quality and graphitic structure of VGN can be obtained from Raman spectroscopy (Figure 3). The presence of G- and G′- bands confirm the graphitic nature of the film. The G- band (~1580 $cm^{-1}$) and G′- bands (~2700 $cm^{-1}$) are due to the in-plane vibration of $sp^2$ carbon ($E_{2g}$ symmetry) and double resonance process, respectively. The shape of G′- bands is the probe for number of graphene layers. The full width at half maximum (FWHM) of G′- bands is found to 67.09±1.32 $cm^{-1}$, which confirms presence of more than five number of graphene layers in each sheet.[13] In addition, the lineshape of G′ fitted with a single Lorentzian peak can be attributed to the turbostratic nature of the film. This result is in good agreement with TEM observation. The D- bands around 1350 $cm^{-1}$ originate from out-of-plane $sp^2$ phonon vibration and D′-band around 1620 $cm^{-1}$ is due to finite crystalline size. The D"-, D+D"-, D+D′- and 2D′- bands are overtone or combination of defect related peaks. The presence of these bands confirms the highly defected nature of the film.[8, 10] The large amount of defects is due to the presence of disorder, arising from large amount of edge states[35], $sp^3$ bonded C–H species, nanographitic base layer and ion-induced defects from the plasma during growth.

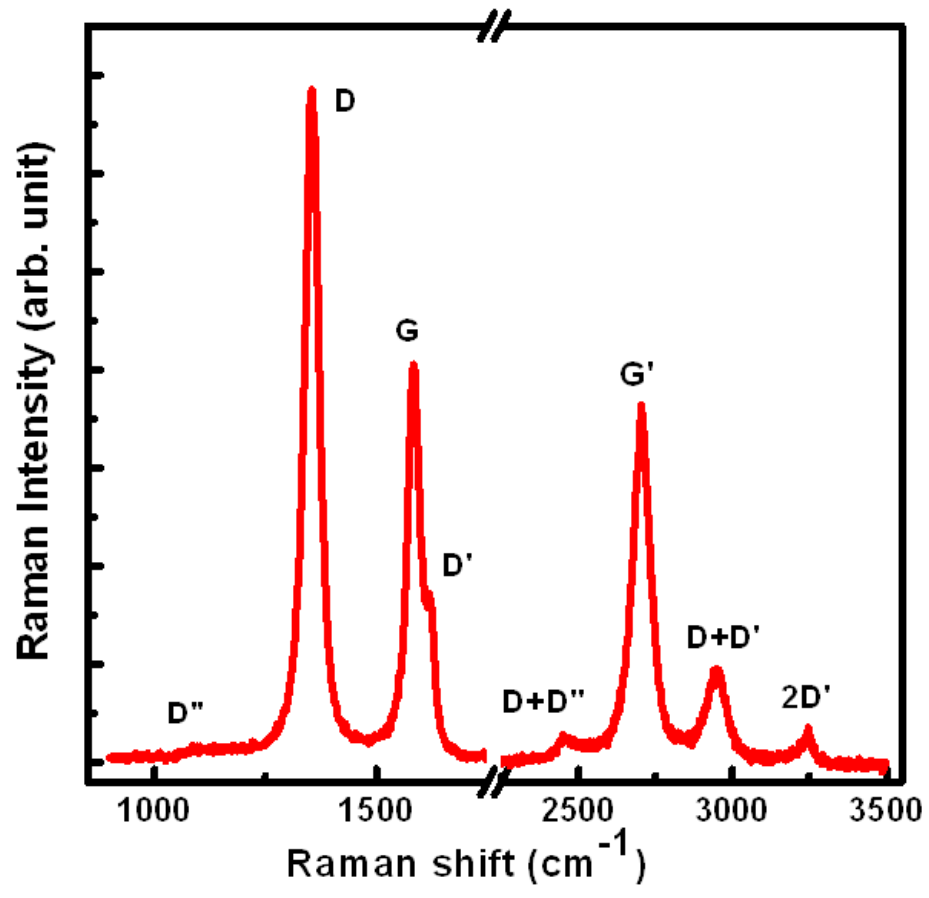

**Figure 3**: Raman spectra of vertical graphene nanosheets

### 3.3. Sheet resistance measurement

The sheet resistance of the VGN is measured by four probe resistive method. The ohmic contact is made by gently placing Ag paste and thin Cu wire. The sheet resistance is found to be 855 Ω/Sq. The high conducting nature of the film decreases the internal resistance thereby enhancing the electron transfer as well as assist in the accumulation of charges for EDL formation.

### 3.4. Electrochemical investigation

The on-chip integration for the micro-electronic device commonly relies on the Si industry.[35] Oakes et al. have shown that Si with porous structure is an excellent candidate for high performance SC.[24] In addition, Si substrate is easily available and cost-effective. Recently, Shredian et al. have shown that directly grown carbon nanospike on n-type Si(100) can be used as alternatives to glassy carbon for potential electrochemical application.[30] Aradilla et al. have shown higher specific capacitance of VGN grown on Si in ionic liquid electrolyte.[1] In view of the above, n-type Si substrate is chosen as substrate to grow VGN and their electrochemical characteristics are studied.

*3.4.1. Cyclic voltammetry*

The electrochemical performance of VGN as an electrode for SCs is investigated by three electrode system in 1M $Na_2SO_4$, 1M KOH, and 1M $H_2SO_4$ electrolytes. Figure 4(a-c) illustrates the respective current response voltammogramms at 100 to 500 mV/s scan rates. The voltage ranges are 0.1-0.6 V, (-)0.2-0.3 V and 0.3-0.8 V, respectively. This implies that the potential window (0.5 V) is same for all the electrolytes studied. It is seen from Figure 4(a-c) that the current response increases with scan rate. The unaltered, mirror symmetric and quasi-rectangular shape of CV at various scan rates for the VGN/$Na_2SO_4$, VGN/KOH and VGN/$H_2SO_4$ systems confirms the near ideal capacitative behavior with good electrochemical reversibility. The areal specific capacitance ($C_s$ in F/cm$^2$) of the electrode/electrolyte systems are calculated from the CV curve using the formula

$$C_s = \frac{\int IdV}{s.A.\Delta V} \quad (1)$$

where, I is the response current, s is the scan rate for the particular CV, A is the electrode area exposed to the electrolyte and ΔV is the potential window. The CV curves for bare Si substrate in different aqueous electrolytes are given in Figure S1 of supporting information. The observation confirms that the result obtained in Figure 4 corresponds to VGN and the contribution from Si substrate is negligible. Here, Si just acts as a mechanical support. The maximum $C_s$ is found for the VGN/KOH system (197 μF/cm$^2$), followed by 188 μF/cm$^2$ for VGN/$H_2SO_4$ system at 100 mV/s. The results are compared and found to be in good agreement with the existing literature.[22] The observed lowest $C_s$ of 44 μF/cm$^2$ for the VGN/$Na_2SO_4$ system is still higher than the double layer capacitance of highly porous interconnected carbon nanosheets and advanced activated carbon in organic electrolyte medium (10 μF/cm$^2$).[29] The present $C_s$ value can be further increased by increasing the height of VGN.[7] This can easily be achieved by optimizing the growth parameters such as growth temperature, plasma power, feedstock gas composition, growth time and substrate to plasma source distance.[6, 11] The differences in $C_s$ values in different systems can be attributed to the difference in ion permeability into the active matrix.[27] The different rate of accessibility of ions into the matrix depends on ionic radius, ionic mobility and molar ionic conductivity.[39] The corresponding parameters are listed in table 2. Though the size of $K^+$ ions are larger compared to $H^+$ ions, the $C_s$ is found to be slightly higher in KOH than that in $H_2SO_4$. This can be attributed to the higher wettability of VGNs in 1MKOH, as shown in Figure 2. The relatively large $C_s$ in $H_2SO_4$ medium can be attributed to the lower cationic radius, higher ionic mobility and molar ionic conductivity of $H^+$ ions. On the other-hand, the relatively less $C_s$ in $Na_2SO_4$ is due to the higher radius of $Na^+$ ions (3.58 A°), lower ionic mobility and molar ionic conductivity compared to those of $H^+$ ions. The $C_s$ values for all the systems are found to decrease with increasing scan rate (Figure 4(d)). This can be attributed to the reduced access of ions to the interior of the interconnected network at higher scan rate.

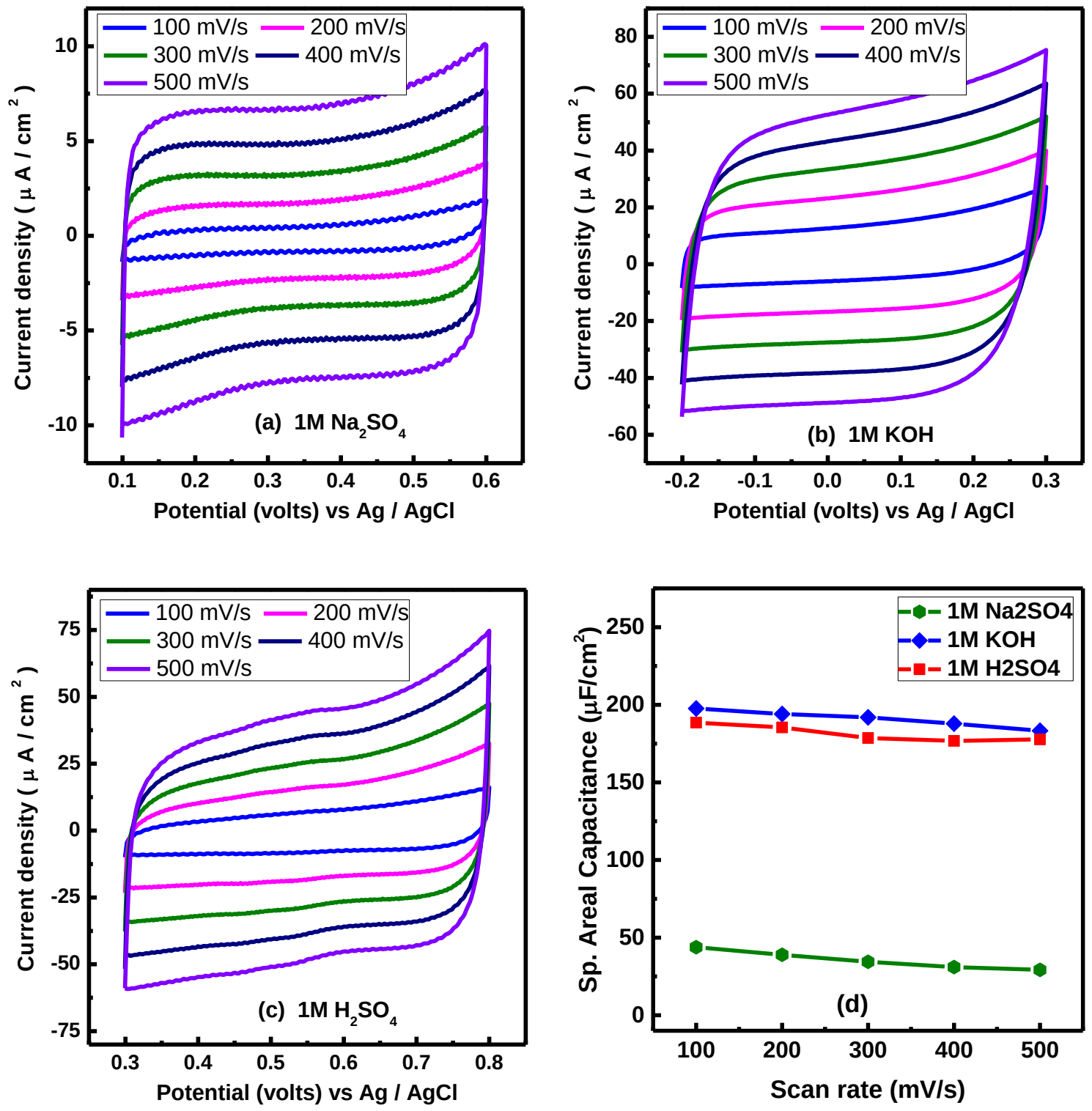

Figure 4: The CV response of VGN in (a) 1M $Na_2SO_4$, (b)1M KOH, and (c)1M $H_2SO_4$ medium at scan rate ranges 100, 200, 300, 400, 500 mV/s. (d) Specific areal capacitance ($C_s$) versus scan rate of VGN under different aqueous electrolytes.

Table. 2 – Characteristic parameters of ions of the distinct aqueous electrolyte[2, 23]

| Ion | Ionic radius (A°) | Radius of hydrated sphere (A°) | Ionic mobility $10^{-8}$ $m^2s^{-1}V^{-1}$ | Molar ionic conductivity $ms.m^2.mol^{-1}$ |
|---|---|---|---|---|
| $Na^+$ | 1.02 | 3.58 | 5.19 | 5.01 |
| $K^+$ | 1.38 | 3.31 | 7.62 | 7.32 |
| $H^+(H_3O^+)$ | 0.021 | 2.80 | 36.23 | 34.96 |
| $OH^-$ | 0.46 | 3.00 | 20.64 | 19.91 |
| $SO_4^{-2}$ | 2.30 | 3.79 | 8.29 | 16.00 |

*3.4.2. Charge discharge*

The charge-discharge responses for different VGN/aqueous electrolyte systems at various current densities are shown in Figure 5(a-c). As it can be seen, the non-linear charging profile is observed in all the electrolytes at lower current density. At higher current density, the voltage–time curves for the VGN/$Na_2SO_4$ and VGN/$H_2SO_4$ systems are nearly linear and symmetric without obvious potential drop. Whereas, the non-linear charging profile and asymmetric voltage-time response is observed for the VGN/KOH system. This indicates better EDL formation across the electrode/electrolyte interface in VGN/$Na_2SO_4$ and VGN/$H_2SO_4$ systems. To get further insights, the energy efficiency (η) is calculated using the equation (2).

$$\eta = \frac{t_D}{t_c} \times 100 \qquad (2)$$

where, $t_D$ and $t_C$ are discharging and charging time. The η of VGN/$Na_2SO_4$, VGN/KOH, and VGN/$H_2SO_4$ systems are 62.1% at 0.76 μA/cm$^2$, 27.6% at 9.09 μA/cm$^2$ and 64.4% at 0.95 μA/cm$^2$, respectively.

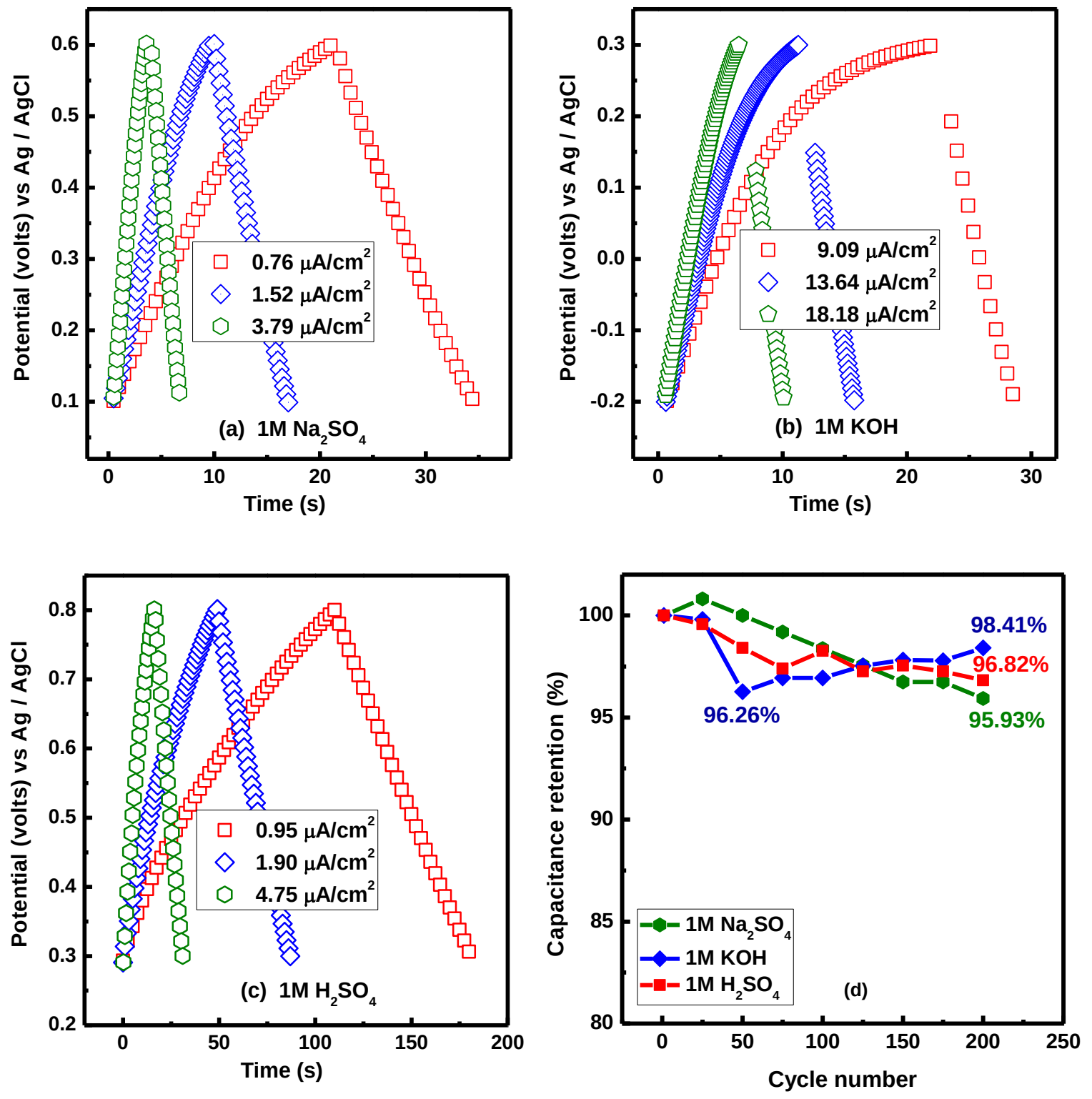

Figure 5: Charge-discharge curve of VGN at different current density under (a) 1M $Na_2SO_4$, (b)1M KOH, and (c)1M $H_2SO_4$ and (d) capacitance retention versus cycle number of VGN in different aqueous electrolytes.

*3.4.3. Cycle stability*

The cycle life is one of the significant parameters of the SC electrode for evaluating the electrochemical performance. The electrochemical stability of the VGN/$Na_2SO_4$, VGN/KOH, and VGN/$H_2SO_4$ systems are evaluated by repeating 200 charge-discharge cycles for corresponding current density and potential window. Figure 5(d) shows the capacitance retention of VGN in various electrolytes as a function of charge-discharge cycle number. The VGN/$H_2SO_4$ system exhibits excellent electrochemical stability (96.82 %) with only 3.18% deterioration of initial $C_s$ after 200 cycles. The VGN/$Na_2SO_4$ system shows 95.93 % capacitance retention. For the VGN/KOH system, the capacity retention shows increasing trends with cycle number after 50 cycles. This can be attributed to the improvement in wettability.[27]

*3.4.4. Electrochemical impedance study*

To find charge transfer kinetics and possible explanation for the difference in electrochemical performance, the electrochemical behavior of the VGN/aqueous electrolyte systems are investigated by EIS.

*Bode plot*

Figure 6(a) represents the variation of phase angle as a function of frequency, which is known as Bode plot. The phase angles are found to be -85°, -65° and -82° at low frequency in the VGN/$Na_2SO_4$, VGN/KOH and VGN/$H_2SO_4$ systems, respectively (Figure 6(a)). In general, phase angle approach to -90° confirms better capacitative performance and rapid charge-discharge process.[9] The present result indicates faster ion diffusion in electrolyte and faster ion adsorption onto the electrode surface in $Na_2SO_4$, and $H_2SO_4$ compared to the KOH. The relaxation time constant, $\tau_0$, defines the time required to deliver the stored charge effectively.[9] To find out $\tau_0$, imaginary part of the capacitance ($C''(\omega)$) is calculated using equation (3) and plotted against frequency for VGN/electrolyte systems, as shown in Figure 6(b).

$$C''(\omega) = \frac{Z'(\omega)}{\omega.[Z'(\omega)^2 + Z''(\omega)^2].s} \qquad (3)$$

The time constant of the systems ($\tau_0$) is obtained from the frequency corresponding to maximum $C''$ ($f_0$), using the relation, $f_0 = 1/\tau_0$. The $\tau_0$ values obtained for the VGN/$Na_2SO_4$,

VGN/KOH and VGN/$H_2SO_4$ systems are 20 ms, 390 ms and 50 ms, respectively. This implies that $Na_2SO_4$ and $H_2SO_4$ electrolytes are better than the KOH electrolyte in the context of quick delivery of stored charge.

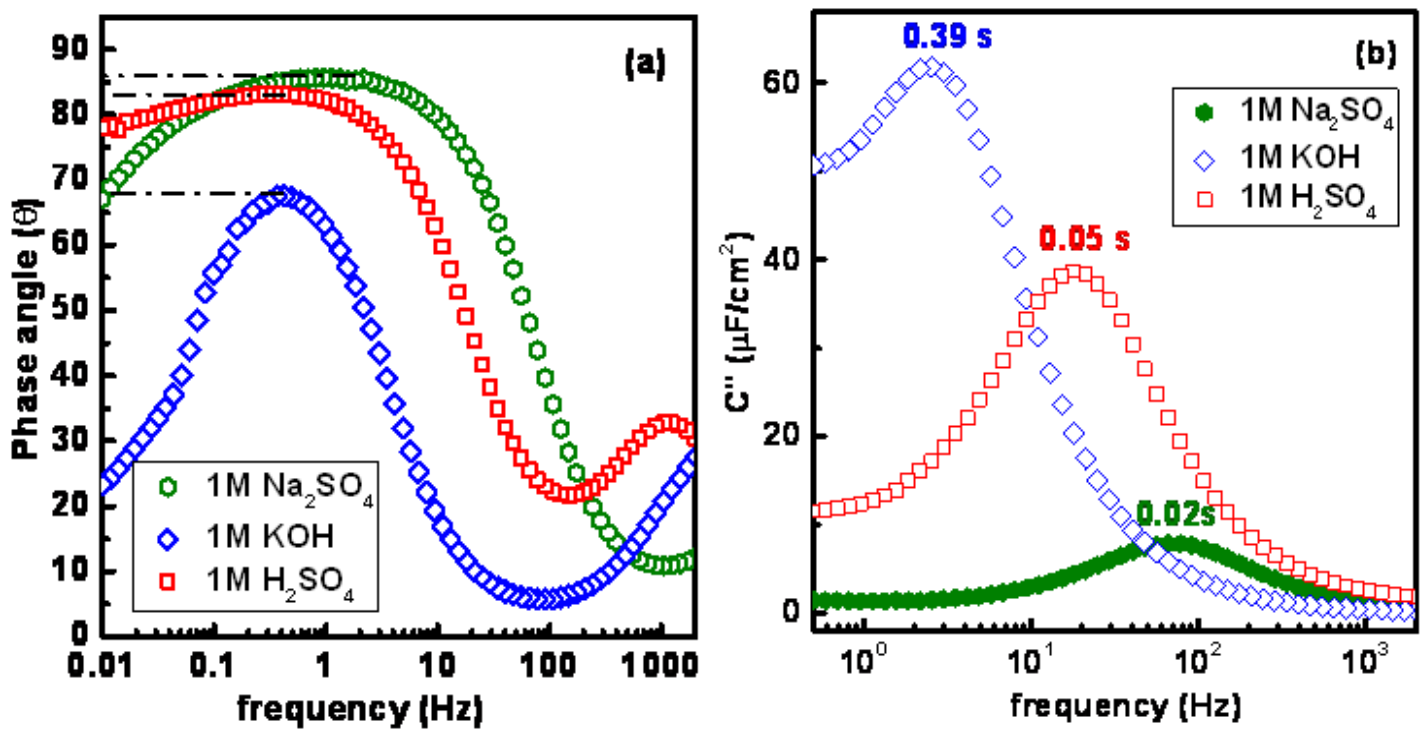

Figure 6: (a) Phase angle versus frequency (Bode plot) and (b) evolution of the imaginary part ofthe complex capacitance versus frequency of VGN under various electrolyte systems

### *3.4.5. Nyquist plot*

The panels (a), (b) and (c) in Figure 7 represents the Nyquist plot of the VGN/aqueous electrolyte systems after 1st and 200$^{th}$ cycles. The insets of Figure 7 show the expanded high frequency region. In general, the impedance spectra consist of three regions: a semicircle at high frequency region, a 45° linear region followed by a vertical line in the intermediate to low frequency regime. At high frequency region (inset of Figure 7(a-c)), compared to VGN/KOH system, a smaller semicircle is observed for the VGN/$Na_2SO_4$ and VGN/$H_2SO_4$ system. This indicates low interfacial charge-transfer resistance. The 45° slopped portion is seen clearly only in the case of VGN/KOH, after 200 cycles of charge-discharge (inset of Figure 7(b)). The 45° slopped portion corresponds to the Warburg region and represents higher variation in ionic diffusion path length and increased obstruction of ion movement.[37] The negligible Warburg region of the VGN/electrolyte system implies short diffusion length of the ions in electrolyte. Additionally, it is seen that after 200 cycle of charge-discharge, the Nyquist plots deviates from the original one for the VGN/KOH system. The observed deviation indicates degradation of the VGN/KOH interface and can be attributed to the oxidation by KOH.[27] The negligible deviation in the spectra after 200 cycles for the VGN/$Na_2SO_4$ and VGN/$H_2SO_4$ systems indicates their excellent electrochemical stability. This observation is in good agreement with those reported by Ruiz et al.[27] The evaluated vertical slopes from Figure 7(a-c) are 87.6°, 84.8°, 86.5° for the VGN/$Na_2SO_4$, VGN/KOH, and VGN/$H_2SO_4$ systems respectively. The slopes close to 90° is indicative of the good capacitative performance.

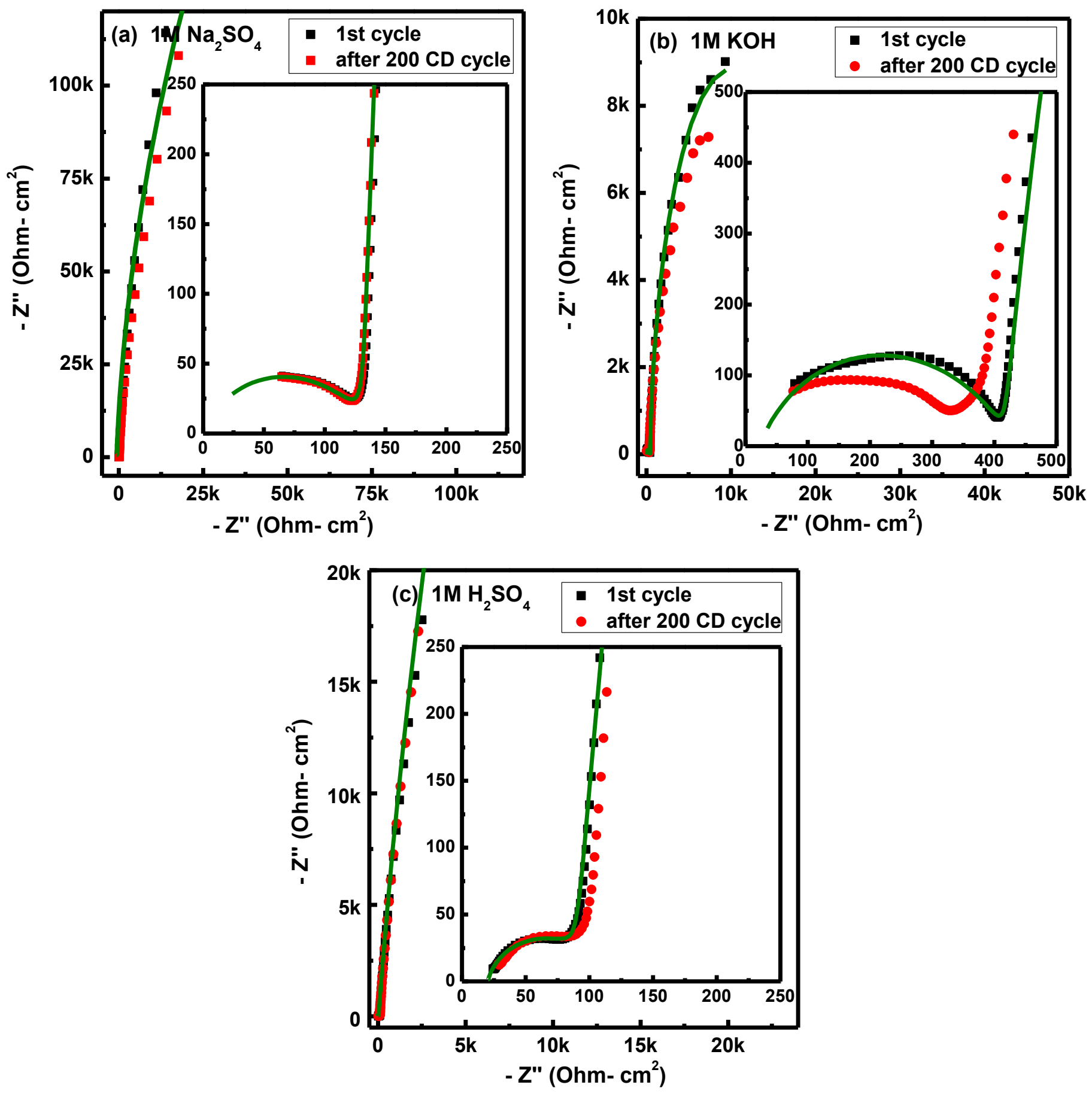

Figure 7: The Nyqust plot of VGN in (a) 1M $Na_2SO_4$, (b)1M KOH and (c)1M $H_2SO_4$ spectra before and after 200 charge-discharge. Inset of panels in (a), (b), (c) represents the expanded high frequency Nyquist plot and solid line (green color) indicates the fitted data using equivalent circuit (Figure 8(b)).

*3.4.6. Simulation of equivalent circuit*

The interpretation of the Nyquist plot usually relies on the construction of a suitable equivalent electric circuit.[3, 15, 16] In general, simple Randles circuit is used to represent the electrochemical system (Figure 8(a)). Our experimental impedance spectra could not be fitted with this simple model, hence modification to the Randles circuit is required. Since, VGN has varying pore size distribution, inhomogeneity and surface disorder, the capacitance can be replaced by constant phase element (CPE). The presence of CPE indicates the frequency dispersion of EDL and can be expressed as

$$Z_{CPE} = {}^{1}\!/_{C'_{dl}.(j\omega)^{n}} \quad ; 0 < n < 1 \qquad (4)$$

where $C'_{dl}$ is proportional to the double layer capacitance and j is the imaginary unit. The exponent (n) is related to the frequency dispersion. For a smooth and clean electrode, n equal to unity represents the ideal capacitative behavior.[3] The physical structure of VGN consists of interconnected array of vertical nanosheets separated by 40-250 nm. The open space provided by the vertical nanosheets, serves as "ion-reservoir" which allows ions migration into the interior of the network. In addition to the interconnected vertical sheets, there are secondary walls normal to them. This gives rise to mesoporous and microporous regions resulting in pore size distribution. The distribution in pore size gives rise to differences in ion mobility and their adsorption. In view of this, a simple modification to the Randles circuit is done to simulate an appropriate equivalent circuit (Figure 8(b)). The fitted curve based on the modified circuit is given in Figure 7(a-c).

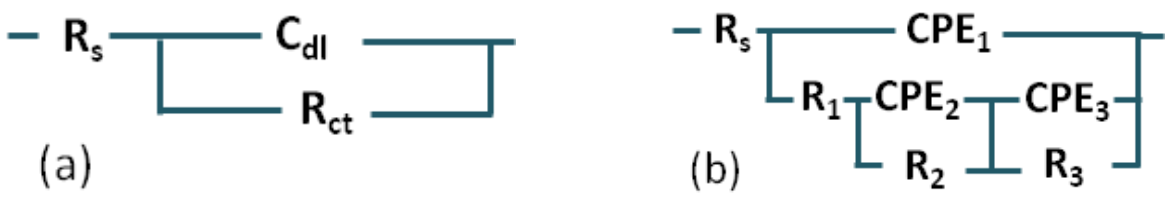

Figure 8: (a) Simple Randles circuit and (b) electrical equivalent circuit for the vertical graphene nanosheets electrode in aqueous medium

In the equivalent circuit, resistor element ($R_s$) is the series resistance of the electrolyte and contact. The $R_s$ value for all systems is found to be almost same. In the case of VGN/electrolyte systems, the formation of EDLC under an ac perturbation involved three processes,[31, 32] viz, (i) charge accumulation on VGN edges in contact with the electrolyte (high frequency region represented by $R_1$, $CPE_1$, $n_1$) (ii) the ion diffusion in the porous channel which allows the migration of ions to the walls of vertical sheets ($R_2$, and ($CPE_2$ and $n_2$)) and (iii) the distributed ionic resistance and double layer capacitance in the micropores and mesopores, nested inside the macropres by the secondary walls, represented by $R_3$, and ($CPE_3$ and $n_3$). The R, CPE and n values extracted from the fitted spectra are given in table 3. The charge transfer resistance ($R_1$) can also be attributed to the potential gradient between the electroactive species in electrolyte and electrode surface.[16] The higher of CPE values in $H_2SO_4$ medium can be attributed to the compatible ionic size with the pore size, higher ionic mobility and higher molar ionic conductivity of $H^+$ ions. The higher CPE values in KOH medium are due to the higher wettability. The higher ionic size, lower ionic mobility and lower ionic molar conductivity resulted in lower CPE values for the VGN/$Na_2SO_4$ system. The total calculated CPE is compared with the capacitance value obtained from CV in table 3. The good agreement between them, further confirms the validity of simulated circuit. The various electrochemical parameters of the VGN/aqueous electrolyte systems are compared in table 4.

Table 3: The extracted parameters of equivalent circuit of VGN under different aqueous electrolytes obtained through EIS spectrum analyser

| | 1M $Na_2SO_4$ | 1M KOH | 1M $H_2SO_4$ |
|---|---|---|---|
| $R_s$ [Ω-cm$^2$] | 21.68 | 20.54 | 21.58 |
| $CPE_1$[μF/cm$^2$] | 1.87 | 1.41 | 14.11 |
| $n_1$ [0<n<1] | 0.7487 | 0.7549 | 0.8375 |
| $R_1$[Ω-cm$^2$] | 110.26 | 370.50 | 72.02 |
| $CPE_2$[μF/cm$^2$] | 17.88 | 143.42 | 90.00 |
| $n_2$[0<n<1] | 0.9761 | 0.9421 | 0.9516 |
| $R_2$[Ω-cm$^2$] | $10^6$ | 19327 | 739460 |
| $CPE_3$[μF/cm$^2$] | 15.39 | 50.00 | 76.35 |
| $n_3$[0<n<1] | 0.2227 | 0.9885 | 0.9746 |
| $R_3$ [Ω-cm$^2$] | 3.13 | 29.40 | 8.07 |
| Total CPE [μF/cm$^2$] | 35.14 | 194.83 | 180.46 |
| $C_s$@100 mV/s [μF/cm$^2$] | 43.84 | 197.63 | 188.48 |

Table 4: Comparison of electrochemical performance of VGN in various aqueous electrolytes

| | 1M $Na_2SO_4$ | 1M KOH | 1M $H_2SO_4$ |
|---|---|---|---|
| *Cs*@100 mV/s [μF/cm$^2$] | 43.84 | 197.63 | 188.48 |
| Capacitance retention after 200 C-D cycles | 95.9 % | 98.4 % | 96.8 % |
| Columbic efficiency | 62.1 % | 27.6 % | 64.4 % |
| Vertical slope from Nyquist plot | 87.6° | 84.8° | 86.5° |
| Phase angle from Bode Plot | -85° | -60° | -80° |
| Time constant (ms) | 20 | 390 | 50 |
| Charge transfer resistance ($R_{ct}$) [Ω-cm$^2$] | 110.26 | 370.50 | 72.02 |

To further investigate the reason behind the different electrochemical behavior, the Raman experiments are carried out on the exposed area of the VGN electrodes after electrochemical performance. Very similar Raman spectra observed before and after electrochemical experiments (Figure S2 in supplementary file), implies the structural stability of VGN. As discussed in section 3.2, the FWHM of Raman modes are a measure of doping and defect, induced after the electrochemical experiment. The very small broadening in FWHM of D peak is observed after the electrochemical studies. The relatively higher broadening observed in KOH can be due to its oxidizing nature. KOH, being an oxidant, is expected to have caused surface oxidation of VGN after 200 cycles of charge-discharge, resulting in local structural damage. This is also reflected in the Nyquist plot. The results are in good agreement with that of Zhu et al.[39] Hence, it is obvious that proper choice of electrolyte solutions is very essential for the development of SCs. The superior capacitance performance, excellent capacity retention, good cycle life and low resistance of VGN/$H_2SO_4$ system can be attributed to the compatible ionic size to access nanoporous structure of VGN efficiently as well as higher ionic conductivity and fast electron kinetics.

## 4. Conclusion

In summary, vertical graphene nanosheets (VGN) are grown directly on n-Si(100) by plasma enhanced chemical vapor deposition. The morphological studies ensure that the VGN have interconnected vertically standing network, where the gap between two sheets varies up to 250 nm. The conductive graphene sheets within the three-dimensional networks serves as a current collector and provides multiple pathways to facilitate the transport of ions and electrons in the electrode surface. The electrochemical performance of VGN is investigated under different aqueous electrolytes namely $Na_2SO_4$, $H_2SO_4$ and KOH. Although VGN/KOH system is capable of slightly higher charge storage than the VGN/$H_2SO_4$ system, the VGN/$H_2SO_4$ system has better performance in terms of capacity retention, fast electron kinetics and minimal change in structure. This can be attributed to the smaller ionic size and higher mobility of $H^+$ ions. The results of this study indicate that not only better conductivity and high surface area of the electrode material, but also the appropriate choice of electrolyte is crucial for the development of next generation supercapacitors.

**Acknowledgement**

One of the authors, Mr. Subrata Ghosh, acknowledge Dept. of Atomic Energy, Govt. of India for Senior Research Fellowship. We acknowledge to Dr. Amrithapandian, Dr. Sandip Dhara and Dr. A. K. Tyagi for their kind support.

# Supplementary information

**Influence of aqueous electrolytes on electrochemical performance of vertical graphene nanosheets supercapacitor electrode**

Subrata Ghosh* [a,b], Bhavana Gupta † [b,c] , Tom Mathews [a,b], A. Das [a,b] and M. Kamruddin * [a,b]

[a]Homi Bhaba National Institute, Mumbai – 400094, India;

[b]Surface and Nanoscience Division, Materials Science Group, Indira Gandhi Centre for Atomic Research, Kalpakkam - 603102, India;

[c]Department of Solar Energy and Environmental Physics, Ben-Gurion University of the Negev, Sede Boqer Campus, Midreshet Ben-Gurion 84990, Israel.

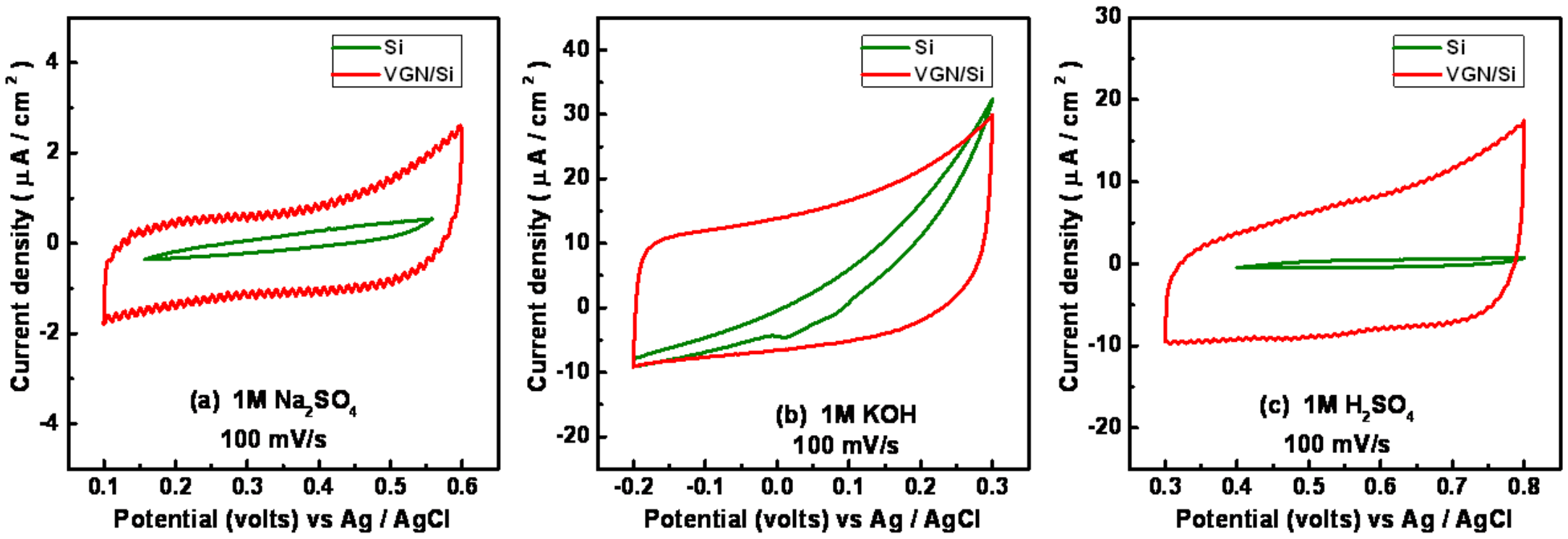

Figure S1. The CV response of Si and VGN on Si under (a) 1M $Na_2SO_4$, (b)1M KOH, and (c)1M $H_2SO_4$ at scan rate 100 mV/s.

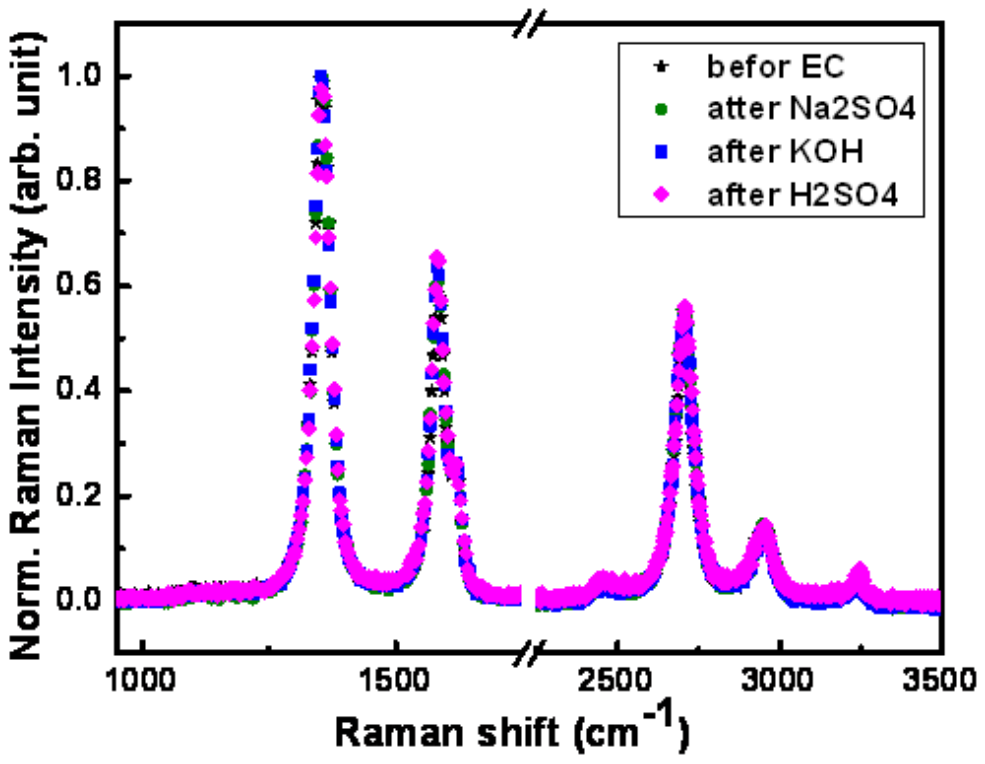

Figure S2. Raman spectra of VGN/Si before and after the electrochemistry experiment under 1M $Na_2SO_4$, 1M KOH, and 1M $H_2SO_4$.